# An Efficient Machine-Learning Approach for PDF Tabulation in Turbulent Combustion Closure

Rishikesh Ranade[a,*], Genong Li[b], Shaoping Li[b], Tarek Echekki[a]

[a]*Department of Mechanical and Aerospace Engineering, North Carolina State University, Raleigh, NC 27695-7910, USA;* [b]*ANSYS INC., Lebanon, New Hampshire 03766, USA*

[*] Corresponding Author. Address: Department of Mechanical and Aerospace Engineering, North Carolina State University, 911 Oval Drive, Campus Box 7910, Engineering Building III, Room 3147, Raleigh, NC 27695-7910, USA. Fax: +1 984 215 8009, E-mail address: rranade@ncsu.edu (R. Ranade).

# An Efficient Machine-Learning Approach for PDF Tabulation in Turbulent Combustion Closure

Probability density function (PDF) based turbulent combustion modelling is limited by the need to store multi-dimensional PDF tables that can take up large amounts of memory. A significant saving in storage can be achieved by using various machine-learning techniques that represent the thermo-chemical quantities of a PDF table using mathematical functions. These functions can be computationally more expensive than the existing interpolation methods used for thermo-chemical quantities. More importantly, the training time can amount to a considerable portion of the simulation time. In this work, we address these issues by introducing an adaptive training algorithm that relies on multi-layer perception (MLP) neural networks for regression and self-organizing maps (SOMs) for clustering data to tabulate using different networks. The algorithm is designed to address both the multi-dimensionality of the PDF table as well as the computational efficiency of the proposed algorithm. SOM clustering divides the PDF table into several parts based on similarities in data. Each cluster of data is trained using an MLP algorithm on simple network architectures to generate 'local' functions for thermo-chemical quantities. The algorithm is validated for the so-called DLR-A turbulent jet diffusion flame using both RANS and LES simulations and the results of the PDF tabulation are compared to the standard linear interpolation method. The comparison yields a very good agreement between the two tabulation techniques and establishes the MLP-SOM approach as a viable method for PDF tabulation.

**Introduction**

The use of a presumed shape probability density function (PDF) is common among a range of turbulent combustion models, including the laminar flamelet approach (Peters, 1983, 1984). In the laminar flamelet approach for non-premixed combustion, this PDF may be parameterized in terms of a mean mixture fraction, $\tilde{Z}$, its variance $\widetilde{Z''^2}$ and the mean dissipation rate, $\tilde{X}$ (Peters, 1983, 1984). This distribution can be used to construct means for thermo-chemical scalars $T$, $\rho$ and $Y_i$, and their source terms. Moreover, these mean quantities can be stored in a PDF table for corresponding $\tilde{Z}$, $\widetilde{Z''^2}$ and $\tilde{X}$. Under non-adiabatic conditions, the Favre-averaged or filtered enthalpy, $\tilde{h}$, provides an additional parameter for the PDF, resulting in the storage of 4D tables for thermo-chemical scalars mean or filtered quantities. Even higher dimensions may be needed to capture either transient effects or an evolving pressure, such as in the case of reciprocating engines.

The *a priori* storage of the thermo-chemical states, $T$, $\rho$ and $Y_i$, means that the equations for species in a CFD code are replaced by a smaller set for the mean mixture fraction, its variance and total enthalpy. The corresponding species' mole fractions are interpolated from the PDF-based table from the solutions of $\tilde{Z}$, $\widetilde{Z''^2}$, $\tilde{X}$ and $\tilde{H}$ at any given point in space and time. Linear interpolations are computationally cheap but may need to store a highly resolved and multi-dimensional PDF table to achieve acceptable accuracy. This can significantly increase the memory storage requirements and constrain the number of parameterized quantities.

Artificial neural networks (ANN) may provide a viable alternative for interpolation for PDF tabulation. ANNs have been applied to many aspects of combustion modeling from reaction rate modeling to chemical mechanism development and reduction (Christo et al., 1996a, b; Blasco et al., 1998, 1999; Chen et al., 2000; Blasco et al., 2000; Ihme et al., 2008, 2009; Sen et al., 2010a, b; Chatzopoulous et al., 2013; Franke et al., 2017; Emami et al., 2012; Ranade et. al., 2019a, 2019b; Ranade and Echekki, 2019a, 2019b). In the present paper, ANNs are used to relate mathematically the parameterized quantities (inputs) and the thermo-chemical quantities (outputs), otherwise stored in a PDF table. Previous studies by Ihme et al. (2009), Emami et al. (2012) and Owoyele et al. (2019) have reported the application of ANNs in a similar context. From these studies, it is remarkably clear that ANN tabulation has an apparent advantage over PDF interpolation in terms of the significant memory savings. However, there are several shortcomings associated with the ANN method that preclude its widespread application:

- ANN training is an important part of this framework since it determines the form and parameters of the mathematical function representing the different scalars included in a multi-dimensional PDF table. It becomes time consuming, especially when these PDF tables store several scalars (~20) over millions of points. As a result, training may account for a substantial portion of the total computation time. Ihme et. al (2009) reported that their ANN takes around 3 days to train 7 scalars in a 3-D PDF table with several million points.

- The mathematical functions resulting from ANN are a combination of nonlinear activation functions (e.g. the tangent sigmoid function) that can be two orders of magnitudes more expensive to compute than linear interpolations. A typical PDF table may store around 20 output variables, which need to be computed at least once per iteration in a CFD run. The number of activation functions computations per iteration can be estimated as follows:

$$N_{tan} = N_l \times N_n \times N_{var}, \qquad (1)$$

where, $N_{tan}$ is the number of *tanh* evaluations, $N_l$ is the number of layers, $N_n$ is the number of neurons per layer and $N_{var}$ is the number of interpolated variables. $N_{tan}$ can be reduced by reducing either $N_l$ or $N_n$ since $N_{var}$ is fixed. Most variables require more than two hidden layer and several neurons per layer to obtain accurate fits, making CFD computation very costly in RANS and LES calculations (Ihme et al., 2009) if ANN-based tabulation is adopted.

- Many combustion applications require PDF tables to store several minor species and pollutants (e.g. NO$_x$, OH). These species are harder to train and require elaborate networks that need to capture multiple peaks, thereby affecting solution accuracy, training and computation costs.

- Finally, the addition of another dimension to the PDF table due to the non-adiabatic nature of the problem adds another level of complexity. This added complexity equally is present with interpolation methods.

These reasons overshadow the memory savings and questions the feasibility of using ANN tabulation over the existing PDF interpolation methods in turbulent combustion simulations, unless strategies to overcome them are not implemented. In this work, we develop an ANN-based framework with an objective to construct small and compact networks that are accurate, as well as, relatively inexpensive in training and computation. Several *a priori* tests are carried out to analyze the performance of this framework. Finally, RANS and LES calculations are carried out for a DLR-A turbulent flame to validate CFD solution accuracy of ANN functional evaluation against the existing PDF interpolations.

**Model Formulation**

In the present work, our strategy to demonstrate the feasibility of ANN-based tabulation is illustrated using the laminar flamelet approach (Peters, 1984). The model is based on the non-adiabatic, steady laminar diffusion flamelet equation in which all the thermo-chemical quantities, $\widetilde{\phi}$, are parameterized by the Favre-averaged quantities for the mixture fraction, $\tilde{Z}$, and its variance, $\widetilde{Z''^2}$, the scalar dissipation rate, $\tilde{X}$, and the total specific enthalpy, $\tilde{h}$. Thus, the Favre-averaged (or filtered for LES) form of the thermo-chemical quantities is parameterized by 4 parameters as follows:

$$\tilde{\phi} = f(\tilde{Z}, \widetilde{Z''^2}, \tilde{X}, \tilde{h}). \qquad (2)$$

A typical PDF table may store approximately 20 thermo-chemical scalars depending on the application and the complexity of the chemical mechanism adopted. In the CFD code, the governing equations are solved for $\tilde{Z}$, $\widetilde{Z''^2}$ and $\tilde{h}$ while $\tilde{X}$ is recovered from a turbulence closure model. The thermo-chemical states corresponding to these parameters either are interpolated from the stored table using linear interpolations or computed using the functional form provided by the ANN training.

The transport equations solved for $\tilde{Z}$, $\widetilde{Z''^2}$ and $\tilde{h}$ and the turbulence closure model for $\tilde{X}$ used in the RANS ($k$ - $\epsilon$ model) formulation in Fluent 19.0 are shown below.

$$\frac{\partial}{\partial t}(\bar{\rho}\tilde{Z}) + \nabla.(\vec{v}\,\bar{\rho}\tilde{Z}) = \nabla\left[\left(\frac{K}{C_p} + \frac{\mu_t}{\sigma_t}\right)\nabla\tilde{Z}\right], \qquad (3)$$

$$\frac{\partial}{\partial t}(\bar{\rho}\,\widetilde{Z''^2}) + \nabla.(\vec{v}\,\bar{\rho}\widetilde{Z''^2}) = \nabla.\left[\left(\frac{K}{C_p} + \frac{\mu_t}{\sigma_t}\right)\nabla\widetilde{Z''^2}\right] + C_g\mu_t.(\nabla\tilde{Z})^2 - C_d\rho\frac{\epsilon}{k}\widetilde{Z''^2}, \qquad (4)$$

$$\frac{\partial}{\partial t}(\bar{\rho}\tilde{h}) + \nabla.(\vec{v}\,\bar{\rho}\tilde{h}) = \nabla.\left(\frac{K_t}{C_p}\nabla\tilde{h}\right) + S_h, \qquad (5)$$

$$\tilde{X} = C_X\frac{\epsilon}{k}\widetilde{Z''^2}. \qquad (6)$$

Here, $K$ is the molecular thermal conductivity, $C_p$ is the mixture specific heat, $\sigma_t$ is the Prandtl number, $\mu$ is the dynamic viscosity, $C_g$, $C_X$ and $C_d$ are model constants. In the case of the LES formulation, the equations for the Favre filtered mixture fraction and the total specific enthalpy remain the same, while Eq. (4) is replaced with an algebraic model for the mixture fraction variance:

$$\widetilde{Z''^2} = C_{var}\,L_s^2\,|\nabla\tilde{Z}|^2, \qquad (7)$$

and Eq. (5) with a different algebraic model for scalar dissipation rate:

$$\tilde{X} = C_X\frac{(\mu_t + \mu)}{\rho\sigma_t}|\nabla\tilde{Z}|^2 \qquad (8)$$

Here, $C_{var}$ is a model constant while $L_s$ is the sub-grid length scale. In each cell, the thermo-chemical states corresponding to these parameters either are interpolated from the stored table using linear interpolations or computed using the functional form provided by ANN training.

*ANN based tabulation method*

ANNs relate input quantities to output quantities through a network of neurons organized within layers: input, hidden and output layers. The input quantities in this case may involve $\widetilde{Z}, \widetilde{Z''^2}, \widetilde{X}$ and $\widetilde{h}$ (in a 4D PDF table) while the output quantities include temperature, density, specific heat, mixture molecular weight and species' mole fractions. The number of neurons in the input layer and output layers of the network is fixed and is equal to the number of input and output quantities, respectively. The number of hidden layers, neurons per hidden layer as well as activation functions per layer may depend on the non-linearity of a problem and the required accuracy needed to capture the output variables. Some of the choices of activation functions include the tangent-sigmoid (hyperbolic tangent), the log-sigmoid (log) or linear functions.

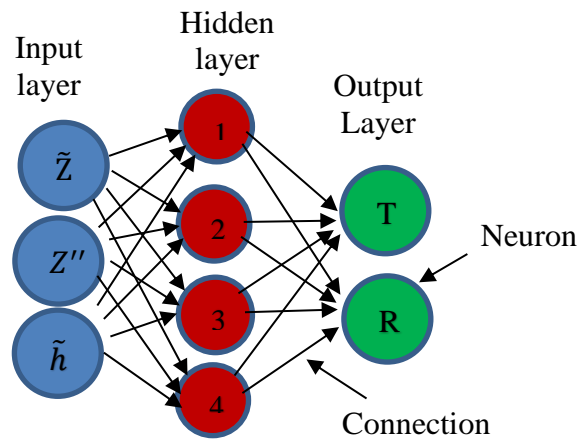

**Figure 1.** Schematic of an ANN with one input layer of 3 neurons, one hidden layer with 4 neurons and an output layer with 2 neurons.

Figure 1 shows a 3-layer network architecture with 3 neurons in the input layer ($\tilde{Z}, \widetilde{Z''^2}, \tilde{h}$), 4 neurons in the hidden layer and 2 neurons in the output layer (*T*, *R*). The evaluation of the outputs based on this architecture is carried out on each cell of the computational domain. In general, the mathematical relationship between input and output variables represented by an ANN architecture with one hidden layer may be expressed as follows:

$$\boldsymbol{\phi_k} = \left[\sum_{i=1}^{N} a_i^k g\left(\sum_{j=1}^{n} b_{ij}^k \bar{x}_j\right) + c_i^k\right], \tag{9}$$

where, $\phi_k = \phi(\tilde{Z}, \widetilde{Z''^2}, \tilde{X}, \tilde{h})$ is the output of the network while $\bar{x}_j$ represents the input vector; *g* is the activation function and $a_i^k, c_i^k, b_{ij}^k$ represent the weights. The goal of ANN training is to determine these weights using a training algorithm. We may see from Eq. (9) that the number of neurons matches the number of activation functions; and, therefore, has a strong effect on the computational cost. Hence, it is beneficial to use a smaller ANN network with less neurons and hidden layers.

Further, ANN training is an important step since it controls, 1) the training time overhead before CFD simulation, 2) the accuracy in predicting thermo-chemical quantities and 3) the computational performance during a CFD simulation. In this work, a multi-layer perceptron (MLP) network is trained using stochastic gradient descent optimization approach with backpropagation for computation of gradients. During training, the output from the network is compared with the desired value for all the training points and the weights are adjusted iteratively until the desired output satisfies a specified tolerance. A common method of error estimation is the mean squared error (MSE). By design, this error estimate gives more importance to larger magnitudes of scalars and may particularly hamper the accuracy of minor/radical species (Bermejo and Cabestany, 2001). Therefore, a stricter definition of error is adopted, which ensures that all the ANN predictions are close to within a specified tolerance of the PDF table value. This definition is prescribed as follows:

$$\tau^k = \tau_a^k + \tau_r \phi_{pdf,j}^k. \tag{10}$$

Here, the net error tolerance, $\tau^k$, for each scalar $k$ is a combination of an absolute error tolerance, $\tau_a$, and a relative error tolerance, $\tau_r$. The goal is to satisfy the net error tolerance, $\tau^k$, at each point in the PDF table. This definition of error ensures that most MLP estimations are within $\tau_r$% of the PDF table values.

*Self-organizing map (SOM) clustering*

In massive, multi-dimensional PDF tables with widely varying thermo-chemical states, determining a smooth function through MLP training can be time-consuming. Therefore, we construct different networks through clustering as a pre-processing step before MLP training. The idea of clustering is to divide the PDF table into different sets based on certain similarities. MLP training is carried out on each set (cluster) to achieve a 'local' mathematical approximation for the thermo-chemical space. The objective here is to attain a) a smaller network, b) an improved accuracy and c) a speed up in training. In this work, we use self-organizing maps (SOM) (Kohonen et al., 1996), which has been successfully implemented within the context of combustion by Franke et al. (2017) in modelling the Sydney flame L, Blasco et al. (2000) for chemistry representation and Ranade and Echekki (2019b) for portioning composition spaces.

In the SOM algorithm (Kohonen et al., 1996), clustering is carried out by tracking the Euclidian distance of each input vector ($\tilde{Z}, \widetilde{Z''^2}, \tilde{X}, \tilde{h}$) of the PDF table and assigning the same cluster to points that are close to one another. Each cluster is assigned a node that is representative of the composition in it. These nodes are arranged in either a 1D or a 2D lattice and SOM creates topologically ordered mappings between input data ($\tilde{Z}, \widetilde{Z''^2}, \tilde{X}, \tilde{h}$) and the nodes of the map. Each node is associated with a weight vector, which is constantly updated during SOM training as and when input data is assigned to it. The weight vectors at all nodes are stored at the completion of training. An input vector $P_i$ of size $N$ is clustered into a cluster $j$ as follows

$$j = argmax\left(-\sqrt{\sum_{i=1}^{N}(W_{1i} - P_i)^2}, -\sqrt{\sum_{i=1}^{N}(W_{2i} - P_i)^2}, \ldots, -\sqrt{\sum_{i=1}^{N}(W_{N_c i} - P_i)^2}\right), \quad (11)$$

where, $W_{1i}, W_{2i}, \ldots, W_{N_c i}$ are weight vectors of the $N_c$ clusters. The input vector, $P_i$ ($\tilde{Z}, Z', \tilde{X}, \tilde{h}$) is assigned to a cluster $j$ whose negative square root of Euclidean distance is maximum. Once the clustering of the data is accomplished, MLP training is carried out on each 'cluster' to find a 'local' MLP regression of thermo-chemical space over that cluster. An advantage of SOM clustering is that it ensures a smooth transition across different clusters (Ranade and Echekki, 2019b). This can minimize the extent of discontinuities occurring across different clusters for most cases. However, a linear blending may be implemented in cases where the discontinuities may be significant. This can be implemented by averaging the MLP-based outputs obtained from neighbouring clusters. SOM clustering may also be used to group multiple reasonably-correlated species thermo-chemical scalars together within the same network for training. Such a strategy is cheaper than developing individual networks for each scalar and more accurate than developing a single network for all scalars. For example, if species like H and O are grouped such that a single network is required to evaluate them, the number of *tanh* (the typical activation function we have used) computations can be substantially reduced in that case. This type of grouping, implemented primarily for species, is carried out using the same SOM-Kohonen (Kohonen et. al, 1996) algorithm. In previous studies (Owoyele et al., 2019) species were grouped based on an affinity function from the correlation matrix of all the species in the PDF table.

*Adaptive MLP-SOM based algorithm*

The objective for MLP-SOM is to build smaller and accurate networks for PDF tabulation. To facilitate this, we have developed an adaptive algorithm to automate the selection of neurons per layer for a specified network architecture to achieve a computationally cheaper CFD and to reduce the training time with strategies of early termination.

The adaptive MLP-SOM algorithm is developed in-house using a C code and is implemented as follows. Starting from a multi-dimensional PDF table, clustering is carried out using the Kohonen SOM algorithm (Kohonen et al., 1996) based on the parameterized quantities in the PDF table. After establishing the clusters and allocating PDF table entries, the data pertaining to each cluster is divided such that a good fraction, say 60%, is reserved for training while the rest is reserved for testing. The MLP network is initialized such that a small number of neurons is specified in the first hidden layer and a fraction of that in subsequent layers. In each MLP training iteration, the updated weights are tested on testing data and training stops when the accuracy specified in Eq. (10) is satisfied or the maximum number of iterations is reached. If the accuracy criterion is not met, the number of neurons in the first hidden layer is incrementally increased and, simultaneously, the neurons of other hidden layers also are updated.

The maximum number of iterations in MLP training can be typically large; and, hence, an early termination strategy is employed if further iterations do not contribute to the convergence of the training. In this work, we are tracking the percentage of points in the PDF table satisfying the tolerance criterion from Eq. (10) in the testing and cross-validation data set to determine the early termination of the training algorithm when a prescribed percentage is achieved after a set number of iterations. An example is to stop if this percentage is greater than 50% after a number of iterations, which is 1/10 of the maximum allowed.

The output of the adaptive algorithm is a set of weights for each output variable in each cluster. These weights are stored in data files or in memory and used for computation during subsequent CFD simulations as opposed to storing an entire PDF table. For example, table 1 compares the memory required for weights based on 4 clusters with a typical 4-D PDF table generated using GRI 3.0 mechanism (Smith et al., 2011) for modelling a DLR-A turbulent jet flame (Meier et al., 2000). A detailed description of the DLR-A has been provided in the following section. The 4-D PDF table is generated using non-adiabatic flamelets with 60 points in $\widetilde{Z}$, 30 points in $\widetilde{Z''^2}$, 30 points in $\widetilde{X}$ and 30 points in $\widetilde{h}$, resulting in 1.62 million points. The PDF table stores temperature, density, specific heat, mixture molecular weight and mole fraction of 20 species, which is a subset of the species in the GRI 3.0 mechanism.

**Table 1.** Memory storage comparison.

| Method | # of hidden layers | Average neurons/layer | Storage (Mb) |
|---|---|---|---|
| MLP-SOM | 2 | 12, 6 | 0.261 |
| MLP-SOM | 3 | 12, 9, 6 | 0.324 |
| MLP-SOM | 4 | 12, 9, 6, 6 | 0.359 |
| MLP-SOM | 5 | 12, 9, 6, 6, 3 | 0.410 |
| PDF | N/A | | 500 |

From Table 1, we can see that the memory requirements for MLP-SOM with different number of hidden layers is more than 1000 times smaller than a conventional PDF table. In general, the memory requirements for MLP-SOM increases proportionally with the number of clusters since weights for all the 'local' approximation functions of all output variables are stored; nonetheless, the overall memory requirement is still considerably lower than that of a PDF table.

Next, we discuss *a priori* tests to analyse the training and computation performance of this algorithm for different network architectures and demonstrate the advantages of clustering in the context of PDF based combustion.

## *A Priori* Validation

### *Need for SOM clustering*

In MLP-SOM based modelling of PDF tables, there is invariably a trade-off between solution accuracy, CFD interpolation time and training time. Massive, multi-dimensional PDF tables with non-linear dependence of thermo-chemical quantities and may require deep MLPs with multiple hidden layers (4-5) or a larger number of neurons per layer to get an acceptable accuracy (Ihme et al., 2009). Although this complexity may provide enough flexibility to model all points in the PDF table, they can also make training and CFD computation very costly. In this section, we have carried out *a priori* tests to explore if clustering truly allows training with simpler network architectures without compromising on accuracy while saving on training and computation time.

We have considered three cases for testing the training and interpolation time, A) MLP training with clustering but without scalar grouping, B) MLP training with clustering and scalar grouping and C) MLP training without clustering or scalar grouping (conventional ANN method), for different network architectures. The adaptive MLP-SOM training algorithm is used in each case to determine the optimum number of neurons per hidden layer for different network architectures. The training accuracy is prescribed (see Eq. (10)) with $\tau_a = \phi_{PDF,i,max} \times 10^{-3}$ and $\tau_r = 4 \times 10^{-2}$, where $\phi_{PDF,i,max}$ is the maximum value of a variable *i* in the PDF table. This tolerance criterion ensures that the resulting MLP-SOM networks in all cases have similar solution accuracies. As long as the training and testing accuracies observed during MLP-SOM training adhere to the tolerance criterion in Eq. (10), the accuracy of the network is independent of the network architecture. Additionally, since the different network architectures exhibit a similar accuracy, it allows a fair comparison of their computational complexity resulting from training and interpolation. All tests have been carried out on the same 4-D PDF table described in the previous section. The PDF data is clustered into 4 clusters using the Kohonen SOM algorithm (Kohonen et al., 1996). Moreover, for case B, the scalars are divided into 9 groups using the same algorithm as shown in Table 2.

**Table 2.** Scalars in different groups using clustering.

| Group # | Scalars |
|---|---|
| 1 | $M_W$, $N_2$ |
| 2 | $\rho$, $O_2$ |
| 3 | H, O, OH, $NH_3$ |
| 4 | CO |
| 5 | $CH_3$, $CH_2O$, $CH_2CO$ |
| 6 | $C_2H_2$, NO, $HO_2$, HCN |
| 7 | T, $H_2O$, $CO_2$ |
| 8 | $C_2H_4$, $C_2H_6$ |
| 9 | $H_2$, $CH_4$ |

Figure 2 shows the time required for training the 24 output thermo-chemical scalars for the three cases using different number of hidden layers. We may observe that the training time for MLP with clustering and scalar grouping (A, B) is less than half of the training time taken for MLP without clustering (C). This corroborates the argument made earlier that SOM based clustering and scalar grouping improves the MLPs ability to fit smooth 'local' functions. The average number of neurons per hidden layer obtained from the adaptive MLP-SOM approach based on the prescribed error is shown in Table 3. Clearly, clustering reduces the complexity of the network, the total number of neurons per network drop considerably.

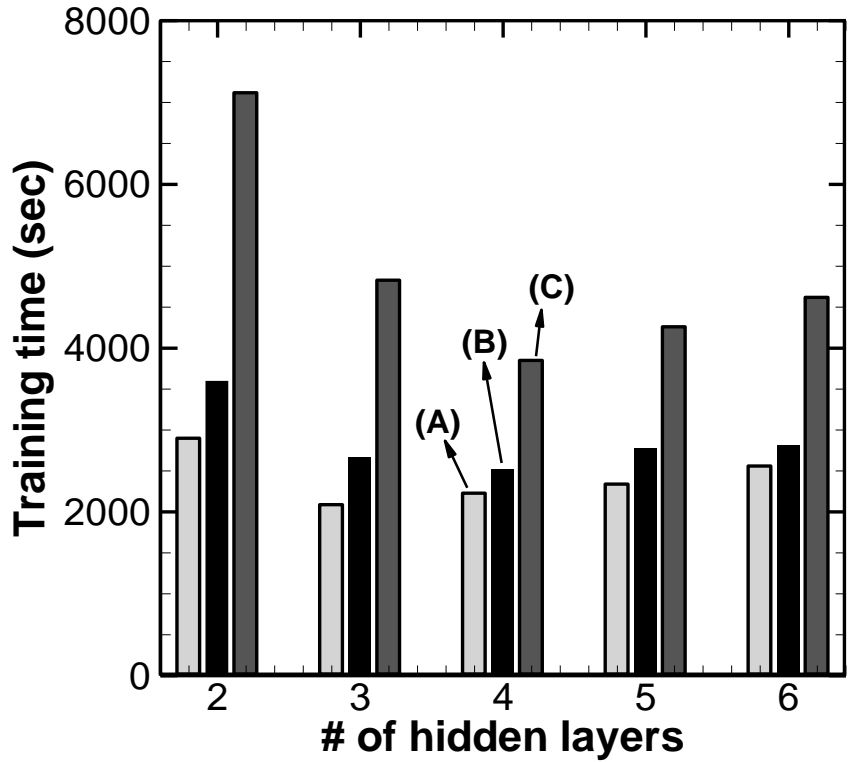

**Figure 2.** Training time needed for various network architectures using (A) MLP-SOM (light grey), (B) MLP-SOM with scalar grouping (black) and (C) MLP without SOM or scalar grouping (dark grey) on a single processor.

**Table 3.** Average neurons/layer observed in A, B and C.

| # of hidden layers | Average neurons/layer (A) | Average neurons/layer (B) | Average neurons/layer (C) |
|---|---|---|---|
| 2 | 12, 6 | 12, 6 | 18, 9 |
| 3 | 12, 9, 6 | 12, 9, 6 | 18, 12, 9 |
| 4 | 12, 9, 6, 6 | 12, 9, 6, 6 | 18, 12, 9, 9 |
| 5 | 12, 9, 6, 6, 3 | 12, 9, 6, 6, 3 | 18, 12, 9, 9, 6 |
| 6 | 12, 9, 6, 6, 3, 3 | 12, 9, 6, 6, 3, 3 | 18, 12, 9, 9, 6, 6 |

For the next test, we have generated 12.96 million points within the bounds of the PDF table and carried out interpolations. Figure 3 shows the interpolation time needed for the various network architectures. The computation time increases with the number of hidden layers. This trend is expected because, in general, the computation time is proportional to the total number of neurons in the network. Moreover, computations using

MLP with clustering or scalar grouping (A, B) are much faster than MLPs without either (C) since clustering reduces neurons per hidden layer and multiple outputs are computed from the same network architecture.

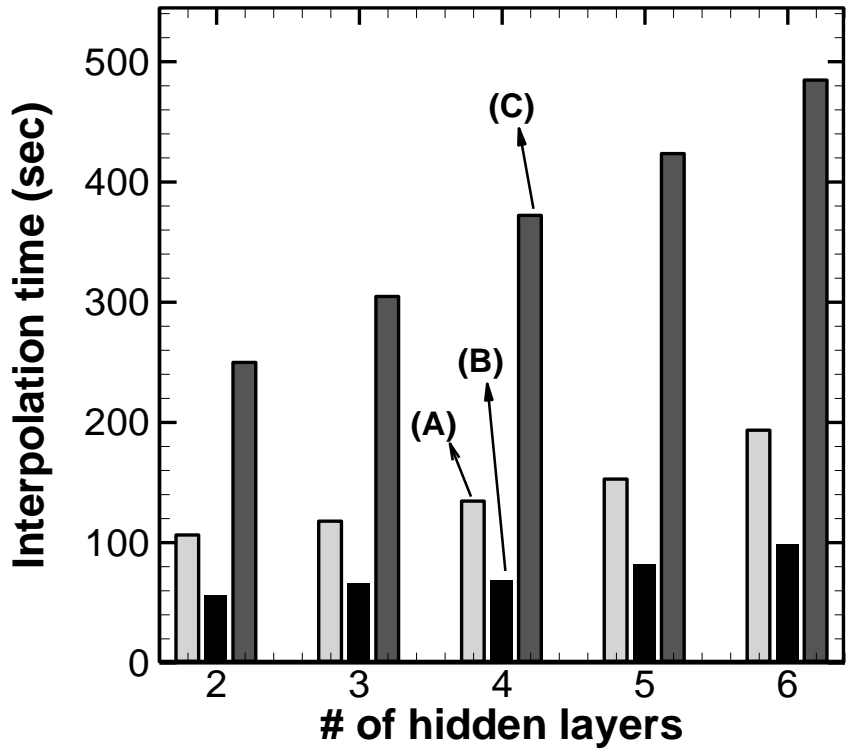

**Figure 3.** Interpolation time needed for various network architectures using (A) MLP-SOM (light grey), (B) MLP-SOM with scalar grouping (black) and (C) MLP without SOM or scalar grouping (dark grey).

The results presented in Figs. 2 and 3 clearly demonstrate the advantages of clustering. Moreover, we have observed that the training time in case of SOM clustering can be significantly reduce by training for these different clusters on different processors. In fact, the ability to run a 'perfectly' parallel problem on many processors provides another crucial advantage to the MLP-SOM approach over MLP-without SOM. The enhancement in training time observed in Fig. 4, which shows the training time vs. the number of hidden layers, can be attributed to the independence of all training events.

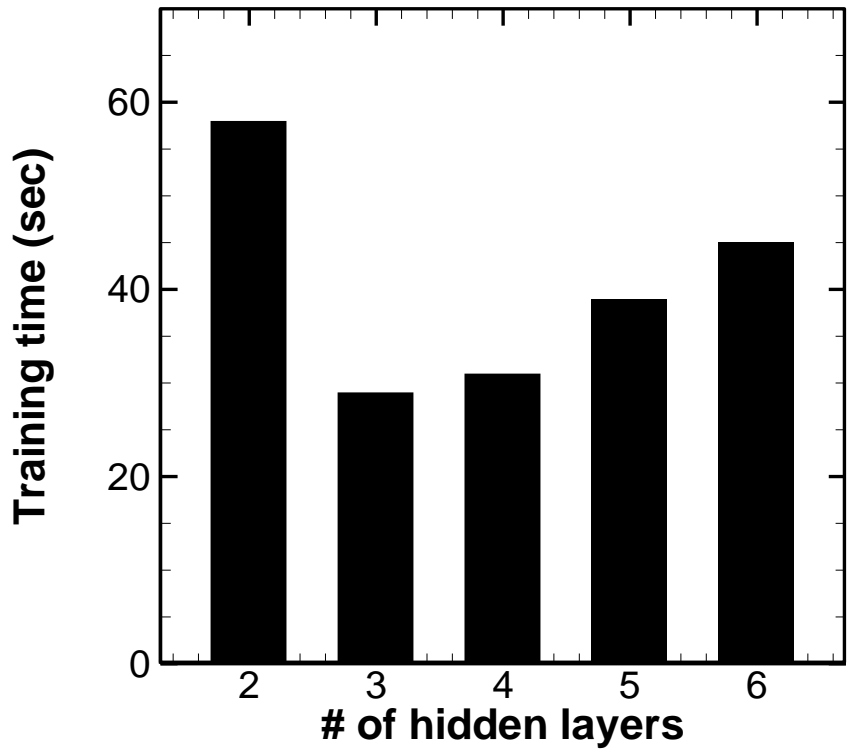

**Figure 4.** Training time needed for various network architectures using MLP-SOM on 96 processors.

*Analysis of the number of SOM clusters*

The choice of the number of clusters in SOM is arbitrary. However, too many clusters can result in fewer data points and affect the solution accuracy as well as training time; while, too few clusters results in more complex network architectures. In this section, we demonstrate this trade-off and explore strategies to choose the optimum number of clusters. The results demonstrated here are based on the same 4-D PDF table described previously. Six tests are carried out for MLP training with 1, 4, 8, 15, 20 and 30 clusters respectively using SOM clustering.

Figure 5 compares the training time using different number of clusters. It may be observed that the training time is comparable for up to 15 clusters but increases dramatically after that. In the case of 30 clusters, there are fewer data points in certain clusters thereby making it harder for the training algorithm to converge. Next, we compare the root mean square error (RMSE) for these different cases by interpolating over 2.56 million points generated randomly within the bounds of the 4-D PDF table. This comparison is given in Table 4 for temperature and 6 species, including $H_2$, $H_2O$, $CO_2$, OH and NO.

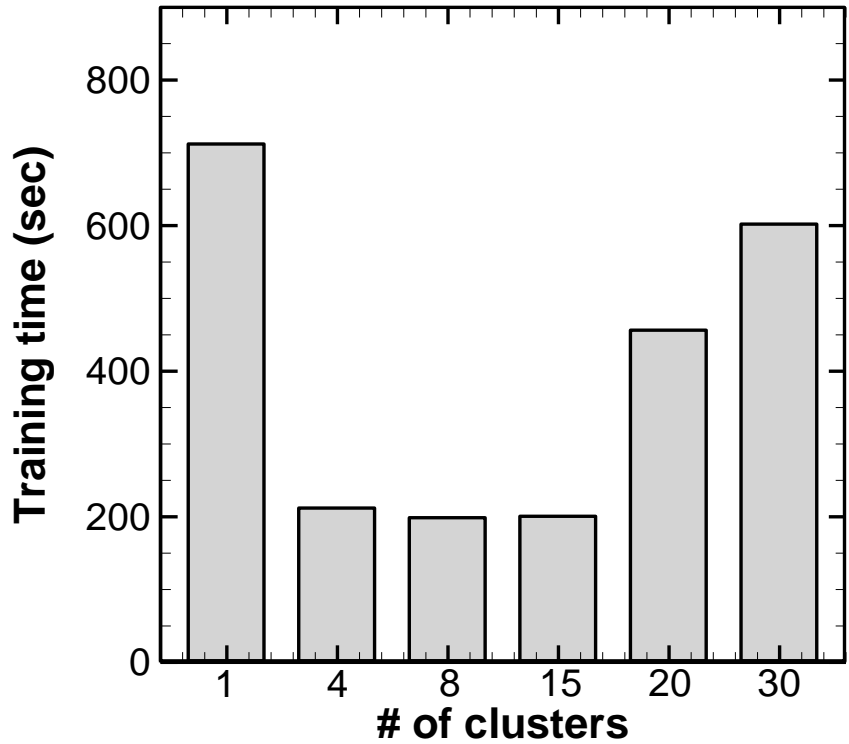

**Figure 5.** Training time needed for different number of clusters using MLP-SOM on 24 processors.

**Table 4.** RMSE for different scalars with number of clusters.

| Scalar | 1 cluster | 4 clusters | 8 clusters | 15 clusters | 20 clusters | 30 clusters |
|---|---|---|---|---|---|---|
| T | 17.82 | 19.43 | 23.86 | 23.15 | 47.26 | 65.85 |
| $H_2$ | 0.0012 | 0.0017 | 0.0022 | 0.0022 | 0.0039 | 0.0065 |
| $H_2O$ | 0.0018 | 0.0019 | 0.0020 | 0.0025 | 0.0045 | 0.0067 |
| $CO_2$ | $3.61 \times 10^{-4}$ | $3.66 \times 10^{-4}$ | $4.13 \times 10^{-4}$ | $4.18 \times 10^{-4}$ | $6.25 \times 10^{-4}$ | $8.05 \times 10^{-4}$ |
| CO | $2.58 \times 10^{-4}$ | $3.11 \times 10^{-4}$ | $3.31 \times 10^{-4}$ | $3.57 \times 10^{-4}$ | $4.51 \times 10^{-4}$ | $6.81 \times 10^{-4}$ |
| OH | $2.53 \times 10^{-5}$ | $2.13 \times 10^{-5}$ | $2.03 \times 10^{-5}$ | $2.32 \times 10^{-5}$ | $8.92 \times 10^{-5}$ | $1.75 \times 10^{-4}$ |
| NO | $1.08 \times 10^{-6}$ | $1.23 \times 10^{-6}$ | $1.14 \times 10^{-6}$ | $1.39 \times 10^{-6}$ | $1.67 \times 10^{-6}$ | $1.87 \times 10^{-6}$ |

We can see in Table 4 that the RMSE increases significantly for the case with 30 clusters. As suggested earlier, too many clusters are counter-productive to training. For a PDF table with points ranging in the order of millions, a cluster choice of 4 to 15 is recommended. In case of bigger tables, the number of clusters can be increased proportionally. Nonetheless, it is evident that the MLP-SOM framework provides cheaper training and better computational performance albeit with reasonable accuracy than existing ANN methods used in the context of PDF based turbulent combustion modelling. Next, we would like to validate CFD solution accuracy and computation time of the framework by comparing it against the existing linear PDF interpolation method.

## *A Posteriori* Studies

### *Model Set Up*

We carry out *a posteriori* studies using a turbulent jet flame, DLR-A (Meier et al., 2000) to test the MLP-SOM approach in RANS and LES using Ansys Fluent 19.0. The objective is to validate the solution accuracy and computation time of the MLP-SOM approach against the default optimized linear PDF interpolation method used in Fluent 19.0. DLR-A (Meier et al., 2000) is a well-characterized flame with a round fuel nozzle of 8 mm diameter, surrounded by a co-flow region with an outer diameter of 0.36 m. The fuel jet consists of a mixture of $CH_4/H_2/N_2$ in the ratio 0.221/0.332/0.447 respectively while the co-flow consists of air. The axial jet and co-flow air velocity are 42.2 m/s and 0.3 m/s respectively. The computational domain is $90d \times 90d \times 90d$ in *x*, *y* and *z* directions, respectively, where *d* is the fuel jet diameter. A hexahedral mesh is used to discretize the domain and has around 2 million cell count.

In the present problem, the combustion chemistry is represented using a non-adiabatic steady diffusion flamelet calculated with GRI 3.0 mechanism (Smith et al., 2011) and the same 4-D PDF table used for *a priori* tests is used here. A pressure-based solver with second order schemes is used to solve governing equations for the parameterized variables ($\tilde{Z}$, $\widetilde{Z''^2}$ and $\tilde{h}$), continuity, momentum and turbulence. The turbulence model used in RANS simulations is the k-ε model while Smagorinsky-Lilly (Smagorinsky, 1963) model is used in LES. The thermo-chemical scalars required in the calculation procedure are either interpolated from the table or computed using the MLP-SOM functions, depending on the case being simulated.

The adaptive MLP-SOM algorithm is used to train the PDF table. 4 clusters are used to ensure enough points in each cluster. The training takes only around a minute on 96 processors; and the results from training are used for both RANS and LES. The network architecture averaged over all the clusters obtained from the adaptive algorithm for different output variables of the PDF table is shown in Table 5. It may be observed that SOM clustering allows the use of simple network architectures with very few neurons to simultaneously train multiple scalars of PDF table. As can be observed from Figs. 2 and 3, these compact networks result in significant reduction in training and computational overheads.

**Table 5.** Average ANN architectures for all variables obtained from MLP-SOM.

| PDF table output variables | Average ANN architecture (neurons in hidden layers) |
|---|---|
| $\rho$, $O_2$ | *14-8* |
| $N_2$, $C_P$, *MOL. WT.* | *6-3* |
| H, O, OH, $NH_3$ | *10-5* |
| $H_2$, $CH_4$ | *10-5* |
| T, $CO_2$, $H_2O$ | *14-8* |
| $C_2H_4$, $C_2H_6$, | *12-6* |
| $C_2H_2$, CO, NO, HCN | *11-6* |
| $CH_3$, $CH_2O$, $HO_2$, $CH_2CO$ | *12-6* |

Next, we would like to compare the computational performance of evaluating the mathematical functions generated from the MLP-SOM approach with that of the existing optimized linear interpolation used in Ansys 19.1. We have randomly generated 12.96 million points within the bounds of the 4D PDF table. In this test, thermo-chemical properties like $T$, $C_p$, $\rho$, $M_W$ etc. and a few selected minor species like NO, $NO_2$, OH, CO etc. in the PDF table are interpolated at each point. In a typical CFD simulation, thermo-chemical properties and these selected minor species are called several times during a single iteration while rest of the species are only needed at the end for post-processing.

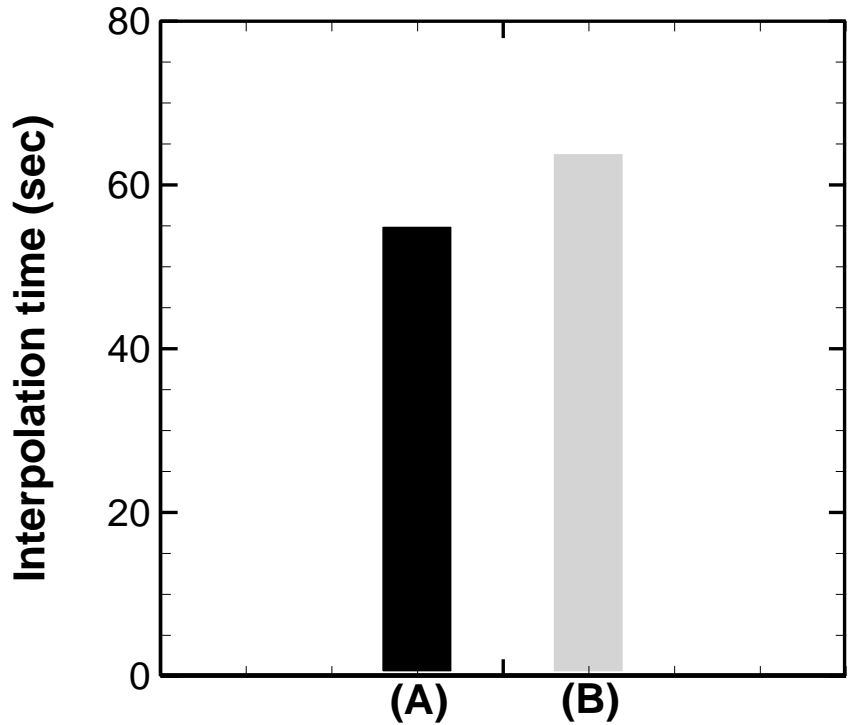

**Figure 6.** Interpolation time comparisons for thermo-chemical scalars using (A) linear interpolation and (B) MLP-SOM function

Figure 6 compares the thermo-chemical property ($T$, $C_p$, $\rho$, $M_W$) and minor species (NO, OH, $NO_2$) interpolation times. Computation times for interpolation of 12.96 million points with both methods are very similar. In the previous work (Ihme et al., 2009) it has been reported that the interpolation of 7 thermo-chemical quantities can take as much as 3-6 times longer than the linear PDF interpolation. A significant reduction in computational time is observed with the MLP-SOM approach since it generates smaller network architectures comprising of fewer neurons. Next, we carry out *a posteriori* studies to test the MLP-SOM approach in CFD and to validate its solution accuracy against the linear PDF interpolation method.

In *a posteriori* analysis, a centerline plot and radial plots at axial locations $x/d$ = 15, 30, 45 and 60 are presented for mean temperature, density, properties and various species mole fractions vs mean mixture fraction (*Z-mean*) for both RANS and LES computations. The idea is to validate the solution provided by the MLP-SOM approach with respect to the linear interpolation and not to gauge which method is more accurate. The goal of this work is merely to present a cheaper and efficient way of representing chemistry. The onus of accurately comparing to experimental data lies solely on the combustion model if the computational solution of MLP-SOM approach can match well with linear PDF interpolations.

*A Posteriori RANS results*

The steady RANS solutions obtained from linear PDF interpolations and MLP-SOM interpolation are presented in Fig. 7. The figure compares radial and axial plots of temperature and the mole fractions of main products of combustion namely, $H_2O$, $CO_2$, $CH_4$, and other intermediate species like CO, $H_2$, H, NO and OH. The results based on the two tabulation methods are in good agreement. The *L*-2 norms showed in Fig. 7 provide a quantitative representation of the difference between the two methods. It may be observed that the errors are very nominal in each case with respect to the maximum value accessed by any thermo-chemical variable. Any small differences may be attributed to the observation that the solutions based on the MLP-SOM interpolation result in smoother plots compared to the linear PDF interpolation. Although not shown here, the remaining minor species stored in the table are captured with a similar accuracy.

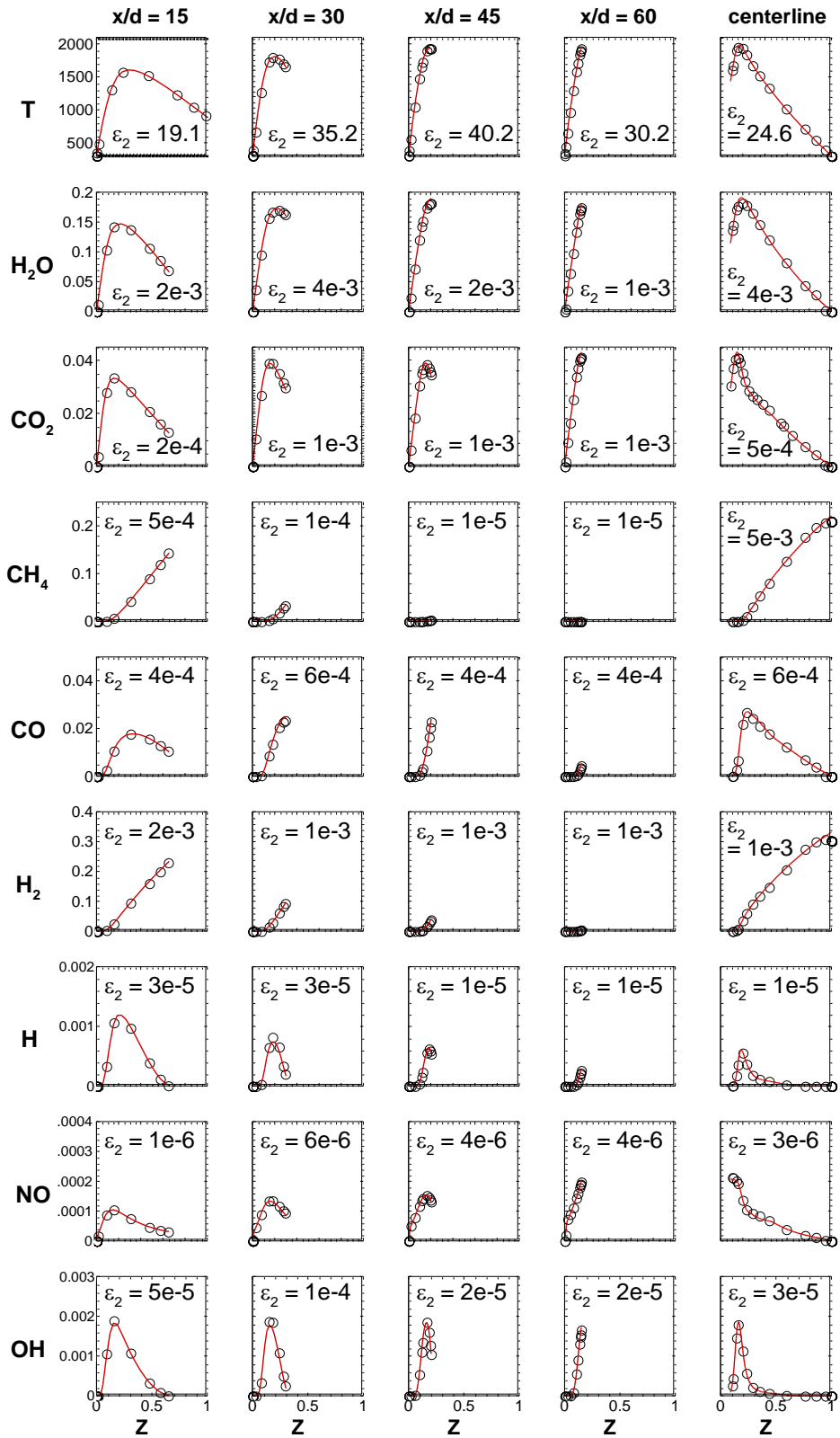

**Figure 7.** Comparison of Steady-state CFD solution for linear PDF interpolation (circle) and MLP-SOM based interpolation (red) at (a) $x/d$ = 15, (b) $x/d$ = 30, (c) $x/d$ = 45, (d) $x/d$ = 60 and (e) centerline. $\epsilon_2$ denotes the $L$-2 norm.

The total CPU times required by the steady RANS solution in Ansys Fluent 19.0 for 11000 iterations on a 1 million-cell mesh is 50 min and 52 minutes using, respectively, the linear interpolation and the MLP-SOM interpolation. The timings based on the two interpolation schemes are comparable and analogous to the results presented in Fig. 6.

*A posteriori LES results*

Long-time error propagation due to functional approximations provided by ANN may result in inaccuracies in large-scale simulations (Ihme et al., 2009). To test the fitness of the MLP-SOM framework in this respect, we have carried out LES calculations of the flame DLR-A using, (i) using linear PDF table interpolation and (ii) MLP-SOM based evaluation. Statistical data is collected over ten flow-through times. LES serves as a stringent test since it runs over several million iterations.

Figures 8 and 9 compare the mean and rms statistics, respectively, of temperature and mass fractions of species including $H_2O$, $CO_2$, CO, OH and NO at various radial and axial locations. The trends observed in mean profiles are similar to the RANS results. Although not shown here, the magnitude of the *L*-2 norm is very similar to the RANS simulation. The rms profiles represent second order moments and are relatively harder to capture. The acceptable agreement of mean and rms profiles after several million iterations in a CFD run provides evidence that the long-time error propagation is negligible and the accuracy of the MLP-SOM based interpolation is satisfactory. The computation time for both cases is comparable; although, the exact computation times are not reported here.

Although not shown here, a significant savings in memory was also obtained by using the MLP-SOM based interpolation. The advantage of machine-learning models in terms of memory saving has been demonstrated extensively in previous works by Ihme et al. (2009), Emami et al. (2012) and Owoyele et al. (2019). The main focus of this work is to present an algorithm that achieves small and compact network architectures and substantially reduce training and interpolation overheads without compromising on solution accuracy.

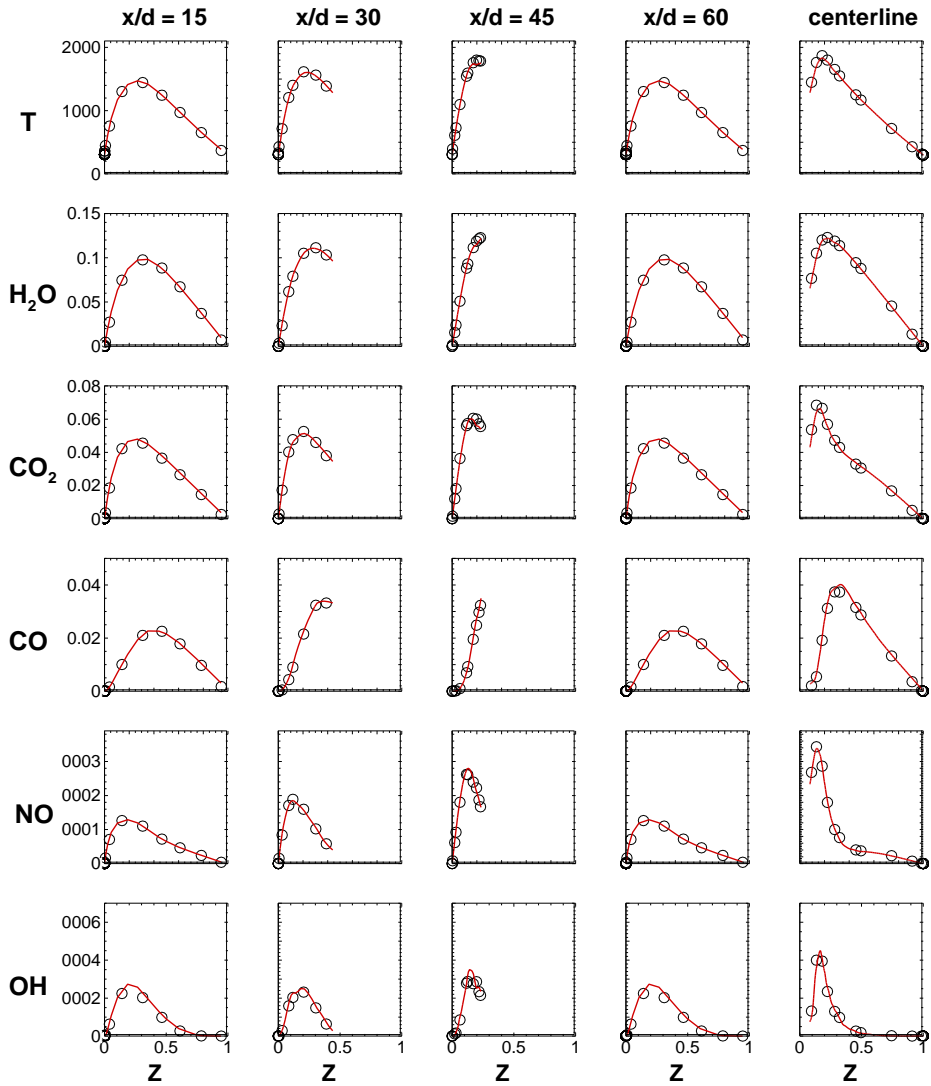

**Figure 8.** Comparison of mean statistics for linear PDF interpolation (symbol) and MLP-SOM based interpolation (line) at (a) *x/d* = 15, (b) *x/d* = 30, (c) *x/d* = 45, (d) *x/d* = 60 and (e) centerline.

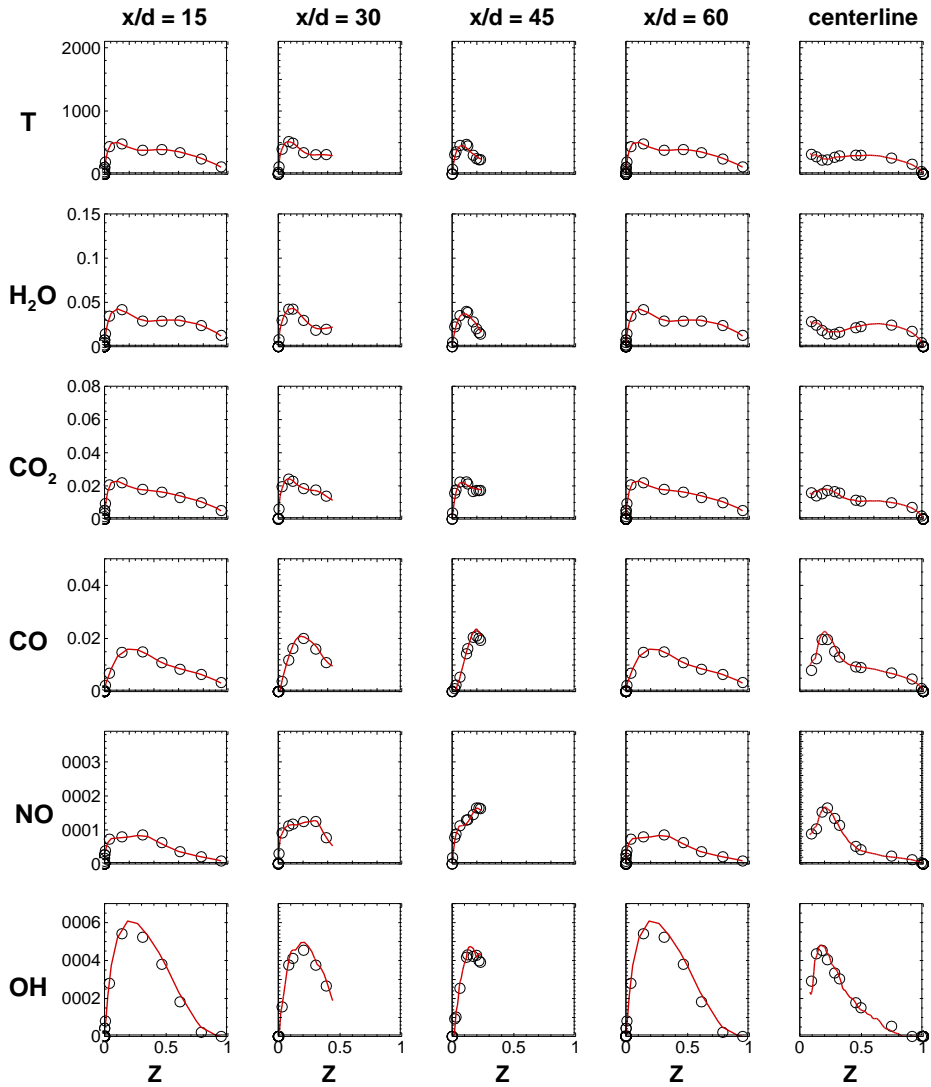

**Figure 9.** Comparison of rms statistics for linear PDF interpolation (symbol) and MLP-SOM based interpolation (line) at (a) *x/d* = 15, (b) *x/d* = 30, (c) *x/d* = 45, (d) *x/d* = 60 and (e) centerline.

**Conclusion**

The main objective of this work is to introduce an adaptive, computationally-efficient MLP-SOM framework to train multi-dimensional PDF tables and explore its feasibility in realistic scenarios where an acceptable solution accuracy is required without expensive computations. Several *a priori* and *a posteriori* tests are carried out to analyse this framework in terms of solution accuracy, computational cost, training cost, memory storage and long-time error propagation during numerical simulations. The following conclusions are drawn from this work:

- The adaptive MLP-SOM algorithm identifies the optimum number of neurons required to fit a thermo-chemical quantity table for a specific network architecture. The simpler networks obtained ensure that the training and interpolation time are significantly reduced and a need for restarting the training is eliminated.
- The training time using SOM based clustering and scalar grouping is around 3 times faster than using a conventional MLP or ANN training methodology. The capability of parallelization over clusters or specie groups provides further speed-up and makes training negligible in comparison to the total simulation time. Moreover, the interpolation time using MLP-SOM based functional evaluation matches well with the linear interpolation method which is a tremendous improvement over previous ANN models in this context (Ihme et al., 2009).
- The RANS and LES studies of flame DLR-A show great agreement between the linear PDF interpolation and MLP-SOM based functional evaluation of thermo-chemical quantities including several minor species. The results at multiple radial and axial locations match well in both cases. Moreover, the CFD solution times are comparable for both methods.
- Finally, the memory requirements of the MLP-SOM method are about 1000 times smaller than those of a PDF table and therefore can easily accommodate a larger number of parameterized quantities.

**Acknowledgements**

Rishikesh Ranade would like to thank the Reacting Flow Development team at Ansys Inc. for financial support and valuable intellectual inputs.